\pdfoutput=1
\documentclass[sigconf]{acmart} 
\usepackage[utf8]{inputenc}
\usepackage[T1]{fontenc}
\usepackage{microtype}
\usepackage{etex}
\usepackage{graphicx}
\usepackage{epstopdf}
\usepackage{array}

\usepackage{booktabs}

\usepackage{mathtools}
\usepackage{comment}
\usepackage{xspace}
\usepackage{xcolor}
\usepackage{paralist}   
\usepackage{balance}

\usepackage{tikz}
\usepackage{pgfplots,pgfplotstable}
\usetikzlibrary{positioning}
\usetikzlibrary{pgfplots.statistics,shapes,arrows,backgrounds}
\usepgfplotslibrary{groupplots}
\pgfplotsset{compat=newest}
\usepackage{adjustbox}
\graphicspath{{./figs/}}

\newcommand{\retecs}{\textsc{Retecs}}
\newcommand{\etal}{et al.}
\newcommand{\secref}[1]{Section~\ref{#1}}

\DeclareUnicodeCharacter{2212}{-}

\begin{document}

\setcopyright{acmcopyright} % Adjust this to match your publishing-rights agreement with ACM.
\acmDOI{10.1145/3092703.3092709}
\acmISBN{978-1-4503-5076-1/17/07}
\acmConference[ISSTA'17]{26th International Symposium  on  Software Testing and Analysis }{July 2017}{Santa Barbara, CA, USA} 
\acmYear{2017}
\copyrightyear{2017}
\acmPrice{15.00}

\begin{CCSXML}
<ccs2012>
<concept>
<concept_id>10011007.10011074.10011099</concept_id>
<concept_desc>Software and its engineering~Software verification and validation</concept_desc>
<concept_significance>500</concept_significance>
</concept>
<concept>
<concept_id>10011007.10011074.10011099.10011102.10011103</concept_id>
<concept_desc>Software and its engineering~Software testing and debugging</concept_desc>
<concept_significance>500</concept_significance>
</concept>
</ccs2012>
\end{CCSXML}
\ccsdesc[500]{Software and its engineering~Software verification and validation}
\ccsdesc[500]{Software and its engineering~Software testing and debugging}

\keywords{Regression testing, Test case prioritization, Test case selection, Reinforcement Learning, Machine Learning, Continuous Integration}

\title[Reinforcement Learning for Automatic Test Case Prioritization and
Selection\\ in Continuous Integration]{Reinforcement Learning for Automatic Test Case Prioritization and Selection in Continuous Integration}

\author{Helge Spieker}
\affiliation{%
  \institution{Simula Research Laboratory}
  \city{Lysaker} 
  \country{Norway} 
}
\email{helge@simula.no}

\author{Arnaud Gotlieb}
\affiliation{%
  \institution{Simula Research Laboratory}
  \city{Lysaker} 
  \country{Norway} 
}
\email{arnaud@simula.no}

\author{Dusica Marijan}
\affiliation{%
  \institution{Simula Research Laboratory}
  \city{Lysaker} 
  \country{Norway} 
}
\email{dusica@simula.no}

\author{Morten Mossige}
\affiliation{%
  \institution{University of Stavanger}
  \city{Stavanger} 
  \country{Norway} 
}
\affiliation{%
  \institution{ABB Robotics}
  \city{Bryne} 
  \country{Norway}
}
\email{morten.mossige@uis.no}

\begin{abstract}
Testing in Continuous Integration (CI) involves test case prioritization, selection, and execution at each cycle.
Selecting the most promising test cases to detect bugs is hard if there are uncertainties on the impact of committed code changes or, if traceability links between code and tests are not available.
This paper introduces \retecs{}, a new method for automatically learning test case selection and prioritization in CI with the goal to minimize the round-trip time between code commits and developer feedback on failed test cases.
The \retecs{} method uses reinforcement learning to select and prioritize test cases according to their duration, previous last execution and failure history.
In a constantly changing environment, where new test cases are created and obsolete test cases are deleted, the \retecs{} method learns to prioritize error-prone test cases higher under guidance of a reward function and by observing previous CI cycles.
By applying \retecs{} on data extracted from three industrial case studies, we show for the first time that reinforcement learning enables fruitful automatic adaptive test case selection and prioritization in CI and regression testing. 
\end{abstract}

%%% Local Variables:
%%% mode: latex
%%% TeX-master: "../SGM17"
%%% End:

\maketitle

\renewcommand{\thefootnote}{}
\footnotetext{This work is supported by the Research Council of Norway (RCN) through the research-based innovation center Certus, under the SFI program.}
\renewcommand{\thefootnote}{\arabic{footnote}}

\section{Introduction}
\label{sec:intro}
\noindent
{\bf Context.} Continuous Integration (CI) is a cost-effective software development practice commonly used in industry \cite{fowler2006continuous,duvall2007continuous} where developers frequently integrate their work. It involves several tasks, including version control, software configuration management, automatic build and regression testing of new software release candidates. Automatic regression testing is a crucial step which aims at detecting defects as early as possible in the process by selecting and executing available and relevant test cases. CI is seen as an essential method for improving software quality while keeping verification costs at a low level \cite{orso2014software,stolberg2009enabling}.

Unlike usual testing methods, testing in CI requires tight control over the selection and prioritization of the most promising test cases. By most promising, we mean test cases that are prone to detect failures early in the process. Admittedly, selecting test cases which execute the most recent code changes is a good strategy in CI, such as, for example in coverage-based test case prioritization \cite{Nardo}.
However, traceability links between code and test cases are not always available or easily accessible when test cases correspond to system tests.
In system testing for example, test cases are designed for testing the overall system instead of simple units of code and instrumenting the system for code coverage monitoring is not easy.
In that case, test case selection and prioritization has to be handled differently and using historical data about failures and successes of test cases has been proposed as an alternative \cite{Kim2002}. Based on the hypothesis that test cases having failed in the past are more likely to fail in the future, \textit{history-based test case prioritization} schedules these test cases first in new CI cycles \cite{Marijan2013}. Testing in CI also means to control the time required to execute a complete cycle. As the durations of test cases strongly vary, not all tests can be executed and \textit{test case selection} is required.

Despite algorithms have been proposed recently \cite{Marijan2013,Noor2015}, we argue that these two aspects of CI testing, namely test case selection and history-based prioritization, can hardly be solved by using only non-adaptive methods. First, the time allocated to test case selection and prioritization in CI is limited as each step of the process is given a contract of time. So, time-effective methods shall be privileged over costly and complex prioritization algorithms. Second, history-based prioritization is not well adapted to changes in the execution environment. More precisely, it is frequent to see some test cases being removed from one cycle to another because they test an obsolete feature of the system.
At the same time, new test cases are introduced to test new or changed features.
Additionally, some test cases are more crucial in certain periods of time, because they test features on which customers focus the most, and then they loose their prevalence because the testing focus has changed. In brief, non-adaptive methods may not be able to spot changes in the importance of some test cases over others because they apply systematic prioritization algorithms.

\noindent
{\bf Reinforcement Learning}.
In order to tame these problems, we propose a new lightweight test case selection and prioritization approach in CI based on reinforcement learning and neural networks.
Reinforcement learning is well-tuned to design an adaptive method capable to learn from its experience of the execution environment.
By adaptive, it is meant, that our method can progressively improve its efficiency from observations of the effects its actions have.
By using a neural network which works on both the selected test cases and the order in which they are executed, the method tends to select and prioritize test cases which have been successfully used to detect faults in previous CI cycles, and to order them so that the most promising ones are executed first.

Unlike other prioritization algorithms, our method is able to adapt to situations where test cases are added to or deleted from a general repository.
It can also adapt to situations where the testing priorities change because of different focus or execution platforms, indicated by changing failure indications.
Finally, as the method is designed to run in a CI cycle, the time it requires is negligible, because it does not need to perform computationally intensive operations during prioritization.
It does not mine in detail code-based repositories or change-logs history to compute a new test case schedule. Instead it facilitates knowledge about test cases which have been the most capable to detect failures in a small sequence of previous CI cycles.
This knowledge to make decisions is updated only after tests are executed from feedback provided by a reward function, the only component in the method initially embedding domain knowledge.

\noindent
The contributions of this paper are threefold:
\begin{enumerate}
\item This paper shows that history-based test case prioritization and selection can be approached as a reinforcement learning problem. By modeling the problem with notions such as states, actions, agents, policy, and reward functions, we demonstrate, as a first contribution, that RL is suitable to automatically prioritize and select test cases;
\item Implementing an online RL method, without any previous training phase, into a Continuous Integration process is shown to be effective to learn how to prioritize test cases. According to our knowledge, this is the first time that RL is applied to test case prioritization and compared with other simple deterministic and random approaches. Comparing two distinct representations (i.e., tableau and neural networks) and three distinct reward functions, our experimental results show that, without any prior knowledge and without any model of the environment, the RL approach is able to learn how to prioritize test cases better than other approaches.
Remarkably, the number of cycles required to improve on other methods corresponds to less than 2-months of data, if there is only one CI cycle per day;   
\item Our experimental results have been computed on industrial data gathered over one year of Continuous Integration. By applying our RL method on this data, we actually show that the method is deployable in industrial settings. This is the third contribution of this paper.
\end{enumerate}

\noindent
{\bf Paper Outline.} The rest of the paper is organized as follows: \secref{sec:notation_and_background} provides notations and definitions.
It also includes a formalization of the problem addressed in our work.
\secref{sec:approach} presents our \retecs{} approach for test case prioritization and selection based on reinforcement learning.
It also introduces basic concepts such as artificial neural network, agent, policy and reward functions.
\secref{sec:exp_evaluation} presents our experimental evaluation of the \retecs{} on industrial data sets, while \secref{sec:related_work} discusses related work.
Finally, \secref{sec:conclusion} summarizes and concludes the paper.

%%% Local Variables:
%%% mode: latex
%%% TeX-master: "../SGM17"
%%% End:
     % 
\section{Formal Definitions}
\label{sec:notation_and_background}
\noindent
This section introduces necessary notations used in the rest of the paper and presents the addressed problem in a formal way. 

\subsection{Notations and Definitions}
\noindent
Let $\mathcal{T}_i$ be a set of test cases $\{ t_1, t_2, \dots, t_N \}$ at a CI cycle $i$. Note that this set can evolve from one cycle to another. Some of these test cases are selected and ordered for execution in a test schedule called $\mathcal{TS}_i$ $(\mathcal{TS}_i \subseteq \mathcal{T}_i)$.
For evaluation purposes, we define further $\mathcal{TS}_i^{total}$ as being the ordered sequence of all test cases ($\mathcal{TS}_i^{total} = \mathcal{T}_i$) as if all test cases are scheduled for execution regardless of any time limit.
Note that $\mathcal{T}_i$ is an unordered set, while $\mathcal{TS}_i$ and $\mathcal{TS}_i^{total}$ are ordered sequences. Following up on this idea, we define a ranking function over the test cases: $rank : \mathcal{TS}_i \rightarrow \mathbb{N}$ where $rank(t)$ is the position of $t$ within $\mathcal{TS}_i$.

In $\mathcal{TS}_i$, each test case $t$ has a verdict $t.verdict_i$ and a duration $t.duration_i$.
Note that these values are only available after executing the test case and that they depend on the cycle in which the test case has been executed.
For the sake of simplicity, the verdict is either $1$ if the test case has passed, or $0$ if it has failed or has not been executed in cycle $i$, i.e. it is not included in $\mathcal{TS}_i$.
The subset of all failed test cases in $\mathcal{TS}_i$ is noted $\mathcal{TS}_i^{fail} = \{t \in \mathcal{TS}_{i} \mbox{ s.t. } t.verdict_i = 0 \}$.
The failure of an executed test case can be due to one or several actual faults in the system under test, and conversely a single fault can be responsible of multiple failed test cases.
For the remainder of this paper, we will focus only on failed test cases (and not actual faults of the system) as the link between actual faults and executed test cases is not explicit in the available data of our context. 
Whereas $t.duration_i$ is the actual duration and only available after executing the test case, $t.duration$ is a simple over-approximation of previous durations and can be used for planning purposes.

Finally, we define $q_i(t)$ as a performance estimation of a test case in the given cycle $i$. By performance, we mean an estimate of its efficiency to detect failures.
The performance $Q_i$ of a test suite $\{t_1,\ldots,t_n\}$ can be estimated with any cumulative function (e.g., sum, max, average, etc.) over $q_i(t_1),\ldots q_i(t_n)$, e.g., $Q_i(\mathcal{TS}_i)=\frac{1}{|\mathcal{TS}_i|}\sum_{t \in \mathcal{TS}_i} q(t)$.

\subsection{Problem Formulation}
\noindent
The goal of any test case prioritization algorithm is to find an optimal ordered sequence of test cases that reveal failures as early as possible in the regression testing process. Formally speaking, following and adapting the notations proposed by Rothermel \etal{} in \cite{Rothermel2001}:
\textit{Test Case Prioritization Problem (TCP)}\\
Let $\mathcal{TS}_i$ be a test suite, 
and $\mathcal{PT}$ be the set of all possible permutations of $\mathcal{TS}_i$, 
let $Q_i$ be the performance, then {\it TCP} aims at finding $\mathcal{TS'}_i$ a permutation of $\mathcal{TS}_i$, such that $Q_i(\mathcal{TS'}_i)$ is maximized.
Said otherwise, TCP aims at finding $\mathcal{TS'}_i$ such that $\forall\;\mathcal{TS}_i \in \mathcal{PT}:\,Q_i(\mathcal{TS'}_i) \geq Q_i(\mathcal{TS}_i)\,.$
Although it is fundamental, this problem formulation does not capture the notion of a time limit for executing the test suite.
\textit{Time-limited Test Case Prioritization} extends the TCP problem by limiting the available time for execution. As a consequence, not all the test cases may be executed when there is a time-contract. Note that other resources (than time) can constrain the test case selection process, too. However, the formulation given below can be adapted without any loss of generality.

\textit{Time-limited Test Case Prioritization Problem (TTCP)}\\
Let $M$ be the maximum time available for test suite execution, then
{\it TTCP} aims at finding a test suite $\mathcal{TS}_i$, such that $Q_i(\mathcal{TS}_i)$ is maximized and the total duration of execution of $\mathcal{TS}_i$ is less than $M$. Said otherwise, TTCP aims at finding $\mathcal{TS}_i$ such that $\forall\;\mathcal{TS'}_i \in \mathcal{PT}:\,Q_i(\mathcal{TS}_i) \geq Q_i(\mathcal{TS'}_i) \land \sum_{t_k \in \mathcal{TS'}_i} t_k.duration \leq M \land \sum_{t_k \in \mathcal{TS}_i} t_k.duration \leq M$.

Still the problem formulation given above does not take into account the history of test suite execution.
In case the links between code changes and test cases are not available as discussed in the introduction, history-based test case prioritization can be used.
The final problem formulation given below corresponds to the problem addressed in this paper and for which a solution based on reinforcement learning is proposed.
In a CI process, TTCP has to be solved in every cycle, but under the additional availability of historical information as a basis for test case prioritization.
{\it Adaptive Test Case Selection Problem (ATCS)}\\
Let $\mathcal{TS}_1,\ldots,\mathcal{TS}_{i-1}$ be a sequence of previously executed test suites, then the {\it Adaptive Test Case Selection Problem} aims at finding $\mathcal{TS}_{i}$, so $Q_{i}(\mathcal{TS}_{i})$ is maximized and $\sum_{t \in \mathcal{TS}_{i}} t.duration \leq M$.

We see that ATCS is an optimization problem which gathers the idea of time-constrained test case prioritization, selection and performance evaluation, without requesting more information than previous test execution results in CI.

%%% Local Variables:
%%% mode: latex
%%% TeX-master: "../SGM17"
%%% End:
       % 
\section{The RETECS method}
\label{sec:approach}
\noindent
This section introduces our approach to the ATCS problem using reinforcement learning (RL), called \textit{Reinforced Test Case Selection} (\retecs{}).
It starts by describing how RL is applied to test case prioritization and selection (\autoref{sec:rl}), then discusses test case scheduling in one CI cycle (\autoref{sec:sc}).
Finally, integration of the method within a CI process is presented (\autoref{sec:ci}).  

\subsection{Reinforcement Learning for Test Case Prioritization}
\label{sec:rl}
In this section, we describe the main elements of reinforcement learning in the context of test case prioritization and selection.
If necessary, a more in-depth introduction can be found in \cite{Sutton1998}.
We apply RL as a {\it model-free} and {\it online learning} method for the ATCS problem.
Each test case is prioritized individually and after all test cases have been prioritized, a schedule is created from the most important test cases, and afterwards executed and evaluated.

\textit{Model-free} means the method has no initial concept of the environment's dynamics and how its actions affect it.
This is appropriate for test case prioritization and selection, as there is no strict model behind the existence of failures within the software system and their detection.

\textit{Online learning} describes a method constantly learning during its runtime.
This is also appropriate for software testing, where indicators for failing test cases can change over time according to the focus of development or variations in the test suite.
Therefore it is necessary to continuously adapt the prioritization method for test cases. 

In RL, an agent interacts with its environment by perceiving its \textit{state} and selecting an appropriate \textit{action}, either from a learned \textit{policy} or by random exploration of possible actions.
As a result, the agent receives feedback in terms of \textit{rewards}, which rate the performance of its previous action.

\begin{figure}
  \centering
  \includegraphics[width=\columnwidth]{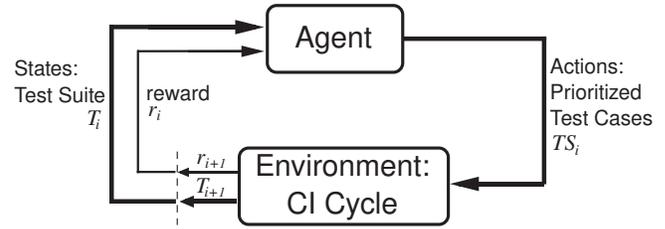}
  \caption{Interaction of Agent and Environment (adapted from \cite[Fig 3.1]{Sutton1998})}
  \label{Fig:rl_process}
\end{figure}

\autoref{Fig:rl_process} illustrates the links between RL and test case prioritization.
A state represents a single test case's metadata, consisting of the test case's approximated duration, the time it was last executed and previous test execution results.
As an action the test case's priority for the current CI cycle is returned.
After all test cases in a test suite are prioritized, the prioritized test suite is scheduled, including a selection of the most important test cases, and submitted for execution.
With the test execution results, i.e., the test verdicts, a reward is calculated and fed back to the agent. 
From this reward, the agent adapts its experience and policy for future actions.
In case of positive rewards previous behavior is encouraged, i.e. reinforced, while in case of negative rewards it is discouraged.
%In terms of the RL domain, prioritization of a full schedule corresponds to an episode with rewards given at the terminal state after execution.

Test verdicts of previous executions have shown to be useful to reveal future failures \cite{Kim2002}.
This raises the question how long the history of test verdicts should be for a reliable indication.
In general, a long history provides more information and allows better knowledge of the failure distribution of the system under test, but it also requires processing more data which might have become irrelevant with previous upgrades of the system as the previously error-prone feature got more stable.
To consider this, the agent has to learn how to time-weight previous test verdicts, which adds further complexity to the learning process.
How the history length affects the performance of our method, is experimentally evaluated in \secref{sec:parameters}.

Of further importance for RL applications are the agent's policy, i.e. the way it decides on actions, the memory representation, i.e. how it stores its experience and policy, and the reward function to provide feedback for adaptation and policy improvement.

In the following, we will discuss these components and their relevance for \retecs{}.

\subsubsection{Reward Functions}
Within the ATCS problem, a good test schedule is defined by the goals of test case selection and prioritization.
It contains those test cases which lead to detection of failures and executes them early to minimize feedback time.
The reward function should reflect these goals and thereby domain knowledge to steer the agent's behavior \cite{Mataric1994}.
Referring to the definition of ATCS, the reward function implements $Q_i$ and evaluates the performance of a test schedule.

Ideally, feedback should be based on common metrics used in test case prioritization and selection, e.g. NAPFD (presented in \autoref{sec:eval_metrics}).
However, these metrics require knowledge about the total number of faults in the system under test or full information on test case verdicts, even for non-executed test cases.
In a CI setting, test case verdicts exist only for executed test cases and information about missed failures is not available.
It is impossible to teach the RL agent about test cases which should have been included, but only to reinforce actions having shown positive effects.
Therefore, in \retecs{}, rewards are either zero or positive, because we cannot automatically detect negative behavior.

In order to teach the agent about both the goal of a task and the way to approach this goal the reward, two types of reward functions can be distinguished.
Either a single reward value is given for the whole test schedule, or, more specifically, one reward value per individual test case.
The former rewards the decisions on all test cases as a group, but the agent does not receive feedback how helpful each particular test case was to detect failures.
The latter resolves this issue by providing more specific feedback, but risks to neglect the prioritization strategy of different priorities for different test cases for the complete schedule as a whole.

Throughout the presentation and evaluation of this paper, we will consider three reward functions.
\begin{definition} \textit{Failure Count Reward}\\
\begin{equation}
reward_i^{\,fail}(t) = |\mathcal{TS}_i^{fail}|\qquad(\forall\,t \in \mathcal{T}_i)
\label{eq:failure_count_reward}
\end{equation}
\end{definition}

In the first reward function \eqref{eq:failure_count_reward} all test cases, both scheduled and unscheduled, receive the number of failed test cases in the schedule as a reward.
It is a basic, but intuitive reward function directly rewarding the RL agent on the goal of maximizing the number of failed test cases.
The reward function acknowledges the prioritized test suite in total, including positive feedback on low priorities for test cases regarded as unimportant.
This risks encouraging low priorities for test cases which would have failed if executed, and could encourage undesired behavior, but at the same time it strengthens the influence all priorities in the test suite have.

\begin{definition} \textit{Test Case Failure Reward}\\
\begin{equation}
reward_i^{\,tcfail}(t) =
\begin{cases}
    1-t.verdict_i  & \quad \text{if } t \in \mathcal{TS}_i\\
    0  & \quad \text{otherwise}\\
\end{cases}
\label{eq:tc_failure_reward}
\end{equation}
\end{definition}

The second reward function \eqref{eq:tc_failure_reward} returns the test case's verdict as each test case's individual reward. Scheduling failing test cases is intended and therefore reinforced. If a test case passed, no specific reward is given as including it neither improved nor reduced the schedule's quality according to available information. Still, the order of test cases is not explicitly included in the reward.
It is implicitly included by encouraging the agent to focus on failing test cases and prioritizing them higher.
For the proposed scheduling method (\autoref{sec:sc}) this automatically leads to an earlier execution.

\begin{definition} \textit{Time-ranked Reward}\\
\begin{equation}
  reward_i^{\,time}(t) = |\mathcal{TS}_i^{fail}| - t.verdict_i \times \sum_{\mathclap{\substack{t_k \in \mathcal{TS}_i^{fail} \land\\ rank(t) < rank(t_k)}}} 1
\label{eq:time_ranked_reward}
\end{equation}
\end{definition}

The third reward function \eqref{eq:time_ranked_reward} explicitly includes the order of test cases and rewards each test case based on its rank in the test schedule and whether it failed. As a good schedule executes failing test cases early, every passed test case reduces the schedule's quality if it precedes a failing test case.
Each test cases is rewarded by the total number of failed test cases, for failed test cases it is the same as reward function \eqref{eq:failure_count_reward}.
For passed test cases, the reward is further decreased by the number of failed test cases ranked after the passed test case to penalize scheduling passing test cases early.

\subsubsection{Action Selection: Prioritizing Test Cases}
Action selection describes how the RL agent processes a test case and decides on a priority for it by using the policy.
The policy is a function from the set of states, i.e., test cases in our context, to the set of actions, i.e., how important each test case is for the current schedule, and describes how the agent interacts with its execution environment.
The policy function is an approximation of the optimal policy.
In the beginning it is a loose approximation, but over time and by gathering experience it adapts towards an optimal policy.

The agent selects those actions from the policy which were most rewarding before.
It relies on its learned experience on good actions for the current state.
Because the agent initially has no concept of its actions' effects, it explores the environment by choosing random actions and observing received rewards on these actions.
How often random actions are selected instead of consulting the policy, is controlled by the exploration rate, a parameter which usually decreases over time.
In the beginning of the process, a high exploration rate encourages experimenting, whereas at a later time exploration is reduced and the agent more strongly relies on its learned policy.
Still, exploration is not disabled, because the agent interacts in a dynamic environment, where the effects of certain actions change and where it is necessary to continuously adapt the policy.
Action selection and the effect of exploration are also influenced by non-stationary rewards, meaning that the same action for the same test case does not always yield the same reward.
  Test cases which are likely to fail, based on previous experiences, do not fail when the software is bug-free, although their failure would be expected.
  The existence of non-stationary rewards has motivated our selection of an online-learning approach, which enables continuous adaptation and should tolerate their occurence.

\subsubsection{Memory Representation}
As noted above, the policy is an approximated function from a state (a test case) to an action (a priority).
There exist a wide variety of function approximators in literature, but for our context we focus on two main approximators.

The first function approximator is the \textit{tableau representation} \cite{Sutton1998}. It consists of two tables to track seen states and selected actions.
In one table it is counted how often each distinct action was chosen per state.
The other table stores the average received reward for these actions.
The policy is then to choose that action with highest expected reward for the current state, which can be directly read from the table.
When receiving rewards, cells for each rewarded combination of states and actions are updated by increasing the counter and calculating the running average of received rewards.

As an exploration method to select random actions, $\epsilon$-greedy exploration is used.
With probability $(1-\epsilon)$ the most promising action according to the policy is selected, otherwise a random action is selected for exploration.

Albeit a straightforward representation, the tableau also restricts the agent.
States and actions have to be discrete sets of limited size as each state/action pair is stored separately.
Furthermore, with many possible states and actions, the policy approximation takes longer to converge towards an optimal policy as more experiences are necessary for the training.
However, for the presented problem and its number of possible states a tableau is still applicable and considered for evaluation.

Overcoming the limitations of the tableau, artificial neural networks (ANN) are commonly used function approximators \cite{VanHasselt2007}.
ANNs can approximate functions with continuous states and actions and are easier to scale to larger state spaces.
The downside of using ANNs are more complex configuration and higher training efforts than for the tableau.
In the context of \retecs{}, an ANN receives a state as input to the network and outputs a single continuous action, which directly resembles the test case's priority.

Exploration is different when using ANNs, too.
Because a continuous action is used, $\epsilon$-greedy exploration is not possible.
Instead, exploration is achieved by adding a random value drawn from a Gaussian distribution to the policy's suggested action.
The variance of the distribution is given by the exploration rate and a higher rate allows for higher deviations from the policy's actions.
The lower the exploration rate is, the closer the action is to the learned policy.

Whereas the agent with tableau representation processes each experience and reward once, an ANN-based agent can be trained differently.
Previously encountered experiences are stored and re-visited during training phase to achieve repeated learning impulses, which is called \textit{experience replay} \cite{Lin1992}.
When rewards are received, each experience, consisting of a test case, action and reward, is stored in a separate replay memory with limited capacity.
If the replay memory capacity is reached, oldest experiences get replaced first.
During training, a batch of experiences is randomly sampled from this memory and used for training the ANN via backpropagation with stochastic gradient descent \cite{zhang2004}.

\subsection{Scheduling}
\label{sec:sc}
Test cases are scheduled under consideration of their priority, their duration and a time limit.
The scheduling method is a modular aspect within \retecs{} and can be selected depending on the environment, e.g. considering execution constraints or scheduling onto multiple test agents.
As an only requirement it has to maximize the total priority within the schedule.
For example, in an environment with only a single test agent and no further constraints, test cases can be selected by descending priority (ties broken randomly) until the time limit is reached.

\subsection{Integration within a CI Process}
\label{sec:ci}
\begin{figure*}
  \centering
  \resizebox{\textwidth}{!}{%\usetikzlibrary{shapes,arrows}

\tikzstyle{block} = [rectangle, draw, 
    text width=3cm, text centered, minimum height=1cm]
\tikzstyle{entity} = [draw,trapezium,trapezium stretches body,trapezium left angle=70,trapezium right angle=-70,minimum height=1cm,text width=2cm, text centered]
\tikzstyle{line} = [draw, -latex']
    
\begin{tikzpicture}[node distance = 3cm, auto]
    % Place nodes
    \node [entity, dashed] (testcases) {Test Cases};
    \node [block, right of=testcases, text width=2cm] (prioritization) {Prioritization};
    \node [entity, right of=prioritization] (priotestcases) {Prioritized\\ Test Cases};
    \node [block, right of=priotestcases, text width=2cm] (scheduling) {Selection \&\\ Scheduling};
    \node [entity, right of=scheduling, text width=2cm] (schedule) {Test Schedule};
    \node [block, right of=schedule, dashed, text width=2.5cm] (execution) {Test Execution};
    \node [entity, below of=execution, dashed, text width=2.5cm, node distance=1.3cm] (devfeedback) {Developer Feedback};
    \node [block, above of=execution, node distance=1.3cm, text width=2.5cm] (evaluation) {Evaluation};
    \node [block, above of=prioritization, node distance=1.3cm, text width=3cm] (policy) {Reinforcement\\ Learning Policy};

    % Draw edges
    \path [line] (testcases) -- (prioritization);
    \path [line] (prioritization) -- (priotestcases);
    \path [line] (priotestcases) -- (scheduling);
    \path [line] (scheduling) -- (schedule);
    \path [line] (schedule) -- (execution);	
    \path [line] (execution) -- (evaluation);
    \path [line] (evaluation) -- (policy);
    \path [line] (execution) -- (devfeedback);
    \path [line] (policy) -- (prioritization);
    \path [line] (policy) -- (prioritization);
\end{tikzpicture}}
  \caption{Testing in CI process: RETECS uses test execution results for learning test case prioritization (solid boxes: Included in RETECS, dashed boxes: Interfaces to the CI environment)}
  \label{fig:ci_cycle}
\end{figure*}
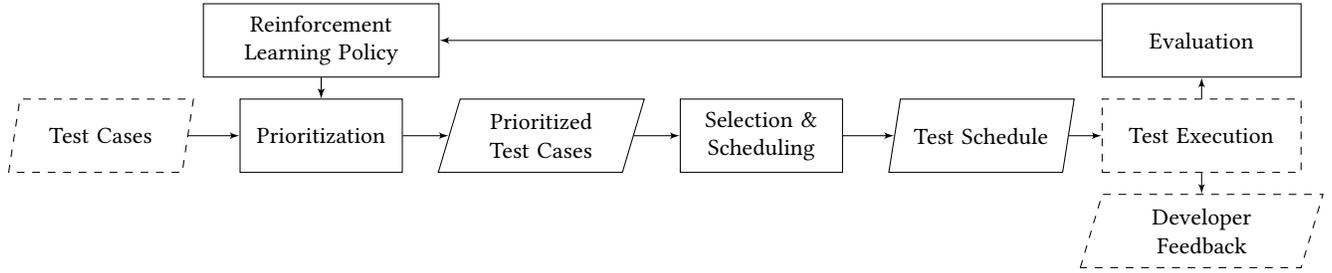

In a typical CI process (as shown in \autoref{fig:ci_cycle}), a set of test cases is first prioritized and based on the prioritization a subset of test cases is selected and scheduled onto the testing system(s) for execution.

The \retecs{} method fits into this scheme by providing the \textit{Prioritization} and \textit{Selection \& Scheduling} steps.
It extends the CI process by requiring an additional feedback channel to receive test results after each cycle, which is the same or part of the information also provided as developer feedback.

%%% Local Variables:
%%% mode: latex
%%% TeX-master: "../SGM17"
%%% End:
	       % 
\section{Experimental Evaluation}
\label{sec:exp_evaluation}
\noindent
In this section we present an experimental evaluation of the \retecs{} method.
During the first part, an overview of evaluation metrics (\autoref{sec:eval_metrics}) is given before the experimental setup is introduced (\autoref{sec:exp_setup}).
In \autoref{sec:results} we present and discuss the experimental results.
A discussion of possible threats (\autoref{sec:threats}) and extensions (\autoref{sec:extensions}) to our work close the evaluation.

Within the evaluation of the \retecs{} method we investigate if it can be successfully applied towards the ATCS problem.
Initially, before evaluating the method on our research questions, we explore how different parameter choices affect the performance of our method.

\begin{itemize}
\item[\textbf{RQ1}] Is the \retecs{} method effective to prioritize and select test cases?
  We evaluate combinations of memory representations and reward functions on three industrial data sets.
\item[\textbf{RQ2}] Can the lightweight and model-free \retecs{} method prioritize test cases comparable to deterministic, domain-specific methods?
  We compare \retecs{} against three comparison methods, one random prioritization strategy and to basic deterministic methods.
\end{itemize}

\subsection{Evaluation Metric} \label{sec:eval_metrics}
In order to compare the performance of different methods, evaluation metrics are required as a common performance indicator.
Following, we introduce Normalized Average Percentage of Faults Detected as the applied evaluation metric.

\begin{comment}
\begin{definition} \textit{Average Percentage of Faults Detected}\\
\begin{equation*}
APFD(\mathcal{TS}_i) = 1 - \frac{\displaystyle\sum_{t \in \mathcal{TS}_i^{fail}} rank(t)}{|\mathcal{TS}_i^{fail}| \times |\mathcal{TS}_i|} + \frac{1}{2 \times |\mathcal{TS}_i|}
\end{equation*}
\label{def:apfd}
\end{definition}
\end{comment}

\begin{definition} \textit{Normalized APFD}\\
\begin{align*}
  NAPFD(\mathcal{TS}_i) =\, &p - \frac{\displaystyle\sum_{t \in \mathcal{TS}_i^{fail}} rank(t)}{|\mathcal{TS}_i^{fail}| \times |\mathcal{TS}_i|} + \frac{p}{2 \times |\mathcal{TS}_i|}\\
  \text{with } &p = \frac{|\mathcal{TS}_i^{fail}|}{|\mathcal{TS}_i^{total,fail}|}\nonumber
\end{align*}
\label{def:napfd}
\end{definition}

Average Percentage of Faults Detected (APFD) was introduced in \cite{rothermel1999} to measure the effectiveness of test case prioritization techniques.
It measures the quality via the ranks of failure-detecting test cases in the test execution order.
As it assumes all detectable faults get detected, APFD is designed for test case prioritization tasks without selecting a subset of test cases.
Normalized APFD (NAPFD) \cite{Qu2007} is an extension of APFD to include the ratio between detected and detectable failures within the test suite, and is thereby suited for test case selection tasks when not all test cases are executed and failures can be undetected.
If all faults are detected ($p = 1$), NAPFD is equal to the original APFD formulation.

\subsection{Experimental Setup} \label{sec:exp_setup}
Two RL agents are evaluated in the experiments.
First uses a tableau representation of discrete states and a fixed number of actions, named \textit{Tableau-based} agent.
And a second, \textit{Network-based} agent uses an artificial neural network as memory representation for continuous states and a continuous action.
The reward function of each agent is not fixed, but varied throughout the experiments.

Test cases are scheduled on a single test agent in descending order of priority until the time limit is reached.

To evaluate the efficiency of the \retecs{} method, we compare it to three basic test case prioritization methods.
First is random test case prioritization as a baseline method, referred to as \textit{Random}.
The other two methods are deterministic.
As a second method, named \textit{Sorting}, test cases are sorted by their recent verdicts with recently failed test cases having higher priority.
For the third comparison method, labeled as \textit{Weighting}, the priority is calculated by a sum of the test case's features as they are used as an input to the RL agent.
Weighting considers the same information as \retecs{} and corresponds to a weighted sum with equal weights and is thereby a naive version of \retecs{} without adaptation.
Although the three comparison methods are basic approaches to test case prioritization, they utilize the same information as provided to our method, and are likely to be encountered in industrial environments.

Due to the online learning properties and the dependence on previous test suite results, evaluation is done by comparing the NAPFD metrics for all subsequent CI cycles of a data set over time.
To account for the influence of randomness within the experimental evaluation, all experiments are repeated 30 times and reported results show the mean, if not stated otherwise.

\retecs{}\footnote{Implementation available at \url{https://bitbucket.org/helges/retecs}} is implemented in Python \cite{Rossum:1995:PRM:869369} using scikit-learn's implementation of artificial neural networks \cite{scikit-learn}.

\subsubsection{Industrial Data Sets}
To determine real-world applicability, industrial data sets from ABB Robotics Norway\footnote{Website: \url{http://new.abb.com/products/robotics}}, \textit{Paint Control} and \textit{IOF/ROL}, for testing complex industrial robots, and \textit{Google Shared Dataset of Test Suite Results} (GSDTSR) \cite{Elbaum2014dataset} are used.\footnote{Data Sets available at \url{https://bitbucket.org/helges/atcs-data}}
These data sets consist of historical information about test executions and their verdicts and each contain data for over 300 CI cycles.

\autoref{tab:industrial_datasets} gives an overview of the data sets' structure.
Both ABB data sets are split into daily intervals, whereas GSDTSR is split into hourly intervals as it originally provides log data of 16 days, which is too short for our evaluation.
Still, the average test suite size per CI cycle in GSDTSR exceeds that in the ABB data sets while having fewer failed test executions. 
For applying \retecs{} constant durations between each CI cycle are not required.

For the CI cycle's time limit, which is not present in the data sets, a fixed percentage of 50\% of the required time is used.
A relative time limit allows better comparison of results between data sets and keeps the difficulty at each CI cycle on a comparable level.
How this percentage affects the results is evaluated in \autoref{sec:sched_time_ratio}.

\begin{table}
  \caption{Industrial Data Sets Overview: All columns show the total amount of data in the data set}
  \centering
  \begin{tabular}{lrrrr}
    \toprule
    Data Set & Test Cases & CI Cycles & Verdicts & Failed\\
    \midrule
    Paint Control & 114 & 312 & 25,594 & 19.36\%\\
    IOF/ROL & 2,086 & 320 & 30,319 & 28.43\%\\
    GSDTSR & 5,555 & 336 & 1,260,617 & 0.25\%\\
    \bottomrule
  \end{tabular}
  \label{tab:industrial_datasets}
\end{table}

\subsubsection{Parameter Selection} \label{sec:parameters}
A couple of parameters allow adjusting the method towards specific environments.
For the experimental evaluation the same set of parameters is used in all experiments, if not stated otherwise.
These parameters are based on values from literature and experimental exploration.

\begin{table}
  \caption{Parameter Overview}
  \centering
  \begin{tabular}{llr}
    \toprule
    RL Agent & Parameter & Value \\
    \midrule
    All & CI cycle's time limit $M$ & $50\% \times \mathcal{T}_i.duration$\\
    & History Length & 4 \\
    \midrule
    Tableau & Number of Actions & 25 \\
    & Exploration Rate $\epsilon$ & 0.2 \\
    \midrule
    Network & Hidden Nodes & 12 \\
    & Replay Memory & 10000 \\
    & Replay Batch Size & 1000 \\
    \bottomrule
  \end{tabular}
  \label{tab:parameters}
\end{table}

\autoref{tab:parameters} gives an overview of the chosen parameters.
The number of actions for the Tableau-based agent is set to 25.
Preliminary tests showed a larger number of actions did not substantially increase the performance.
Similar tests were conducted for the ANN's size, including variations on the number of layers and hidden nodes, but a network larger than a single layer with 12 nodes did not significantly improve performance.

\begin{figure}
  \centering
  \begin{adjustbox}{trim=0.5cm 0.5cm 0.5cm 0.3cm}
  \input{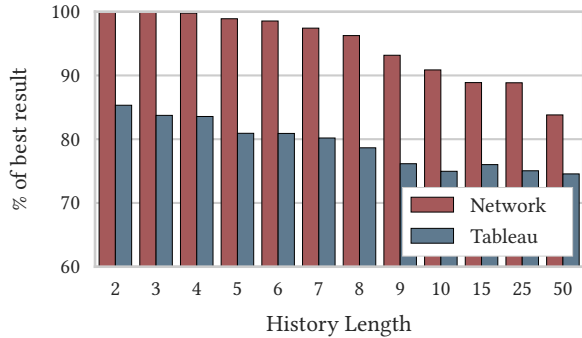}%
  \end{adjustbox}
  \caption{Relative performance of different history lengths. A longer history can reduce the performance due to more complex information. (Data set: ABB Paint Control)}
  \label{fig:history_length}%
\end{figure}

The effect of different history lengths is evaluated experimentally on the Paint Control data set.
As \autoref{fig:history_length} shows, does a longer history not necessarily correspond to better performance.
From an application perspective we interpret the most recent results to also be the most relevant results.
Many historical failures indicate a relevant test case better than many passes, but individual consideration of each of these results on their own is unlikely to lead to better conclusions of future verdicts.
From a technical perspective, this is supported by the fact, that a longer history increases the state space of possible test case representations.
A larger state space is in both memory representations related to a higher complexity and requires generally more data to adapt, because the agent has to learn to handle earlier execution results differently than more recent ones, for example by weighting or aggregating them.

\subsection{Results} \label{sec:results}
\subsubsection{RQ1: Learning Process \& Effectiveness} \label{sec:rq1} % Evaluation of component combinations

\begin{figure*}[!ht]
  \centering
  \begin{adjustbox}{trim=0.7cm 0.5cm 0.5cm 0.35cm}
\resizebox{\textwidth}{!}{\input{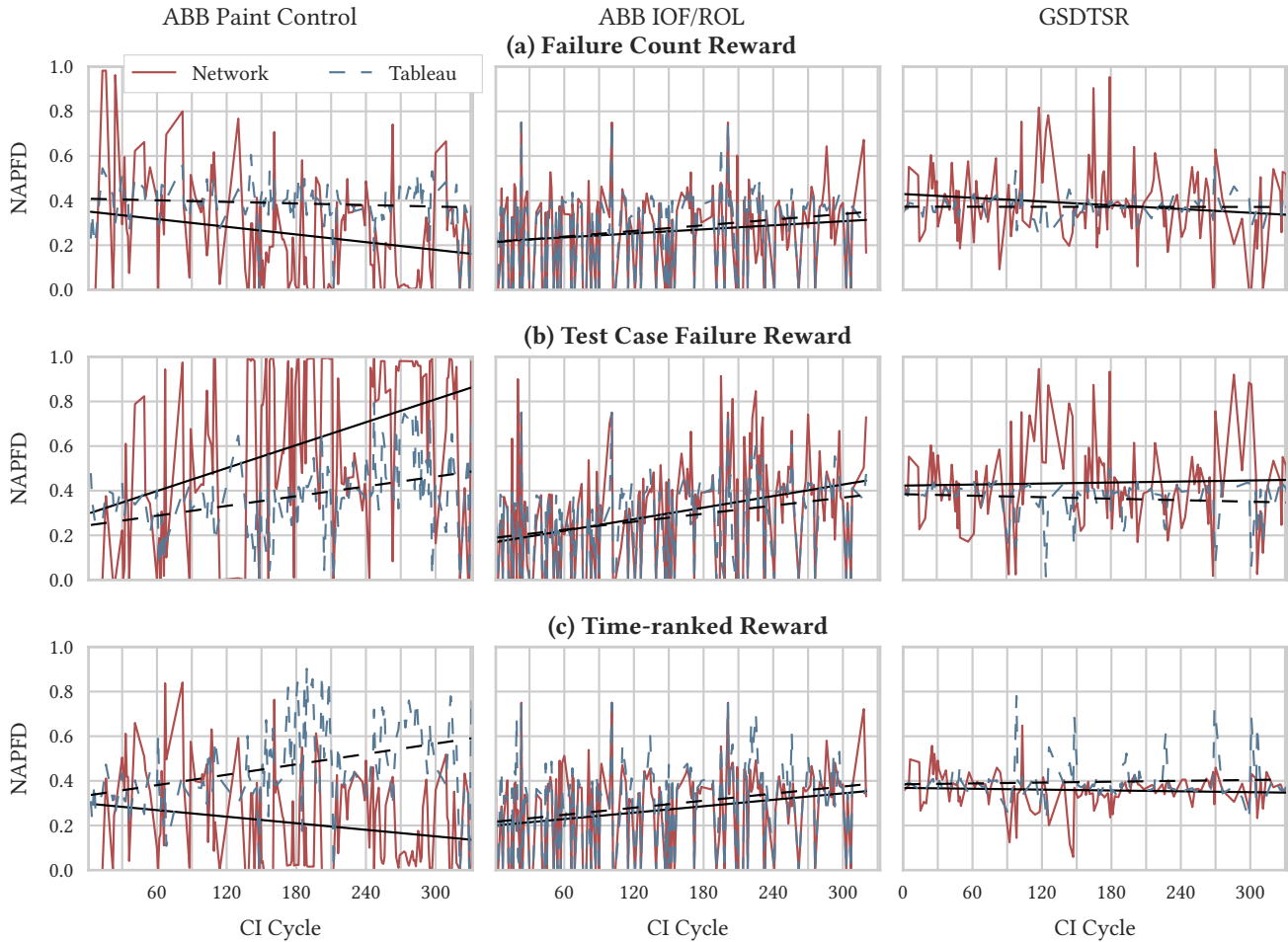}}%
\end{adjustbox}
\caption{Comparison of reward functions and memory representations: A Network-based agent with Test Case Failure reward delivers best performance on all three data sets (Black lines indicate trend over time)}
\label{fig:res_rewardfuns}
\end{figure*}

\autoref{fig:res_rewardfuns} shows the performance of Tableau- and Network-based agents with different reward functions on three industrial data sets.
Each column shows results for one data set, each row for a particular reward function.

It is visible that the combination of memory representation and reward function strongly influences the performance.
In some cases it does not support the learning process and the performance stays at the initial level or even declines.
Some combinations enable the agent to learn which test cases to prioritize higher or lower and to create meaningful test schedules.

Performance on all data sets is best for the Network-based agent with the Test Case Failure reward function.
It benefits from the specific feedback for each test case and learns which test cases are likely to fail.
Because the Network-based agent prioritizes test cases with continuous actions, it adapts more easily than the Tableau-based agent, where only specific actions are rewarded and rewards for one action do not influence close other actions. 

In all results a similar pattern should be visible.
Initially, the agent has no concept of the environment and cannot identify failing test cases, leading to a poor performance.
After a few cycles it received enough feedback by the reward function to make better choices and successively improves.
However, this is not true for all combinations of memory representation and reward function.
One example is the combination of Network-based agent and Test Case Failure reward.
On Paint Control, the performance at early CI cycles is superior to the Tableau-based agent, but it steadily declines due to misleading feedback from the reward function.

One general observation are performance fluctuations over time.
These fluctuations are correlated to noise in the industrial data sets, where failures in the system occur for different reasons and are hard to predict.
For example, in the Paint Control data set between 200 and 250 cycles a performance drop is visible.
For these cycles a larger number of test cases were repeatedly added to the test suite manually.
A large part of these test cases failed, which put additional difficulty on the task.
However, as the test suite was manually adjusted, from a practical perspective it is arguable whether a fully automated prioritization technique is feasible during these cycles.

In GSDTSR only few failed test cases occur in comparison to the high number of successful executions.
This makes it harder for the learning agent to discover a feasible prioritization strategy.
Nevertheless, as the results show, it is possible for the Network-based agent to create effective schedules in a high number of CI cycles, albeit with occasional performance drops.

Regarding RQ1, we conclude that it is possible to apply \retecs{} on the ATCS problem.
In particular, the combination of memory representation and reward function strongly influences the performance of the agent.
We found both Network-based agent and Test Case Failure Reward, as well as Tableau-based agent with Time-ranked Reward, to be suitable combinations, with the former delivering an overall better performance.
The Failure Count Reward function does not support the learning processes of the two agents.
Providing only a single reward value without further distinction is not helping the agents towards an effective prioritization strategy.
It is better to reward each test case's priority individually according to its contribution to the previous schedule.

\subsubsection{RQ2: Comparison to Other Methods} \label{sec:rq2} % Comparison with basic methods
\begin{figure*}
  \centering
  \begin{adjustbox}{trim=0.5cm 0.5cm 0.6cm 0.25cm}
  \scalebox{0.95}{\input{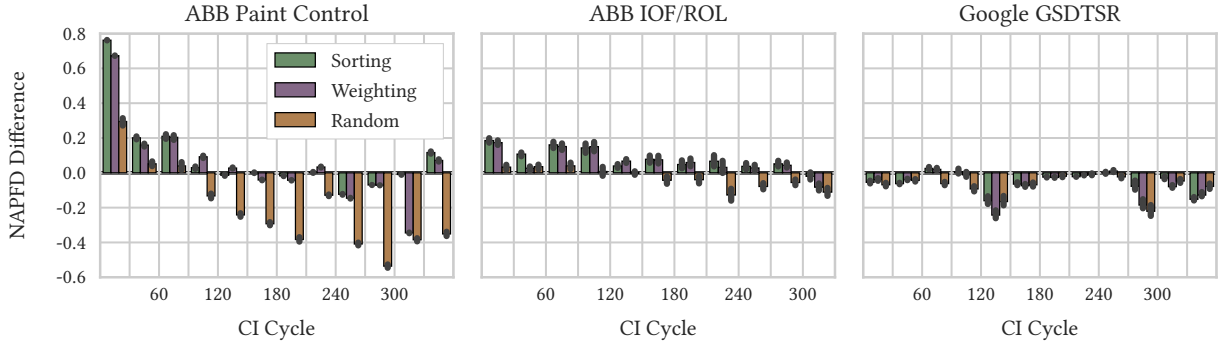}}%
  %\resizebox{\textwidth}{!}{\input{figs/rq3_napfd_bar_abs_grouped.pgf}}%
  \end{adjustbox}
\caption{Performance difference between network-based agent and comparison methods: After an initial exploration phase RETECS adapts to competitive performance. Each group of bars compares 30 CI cycles.}
\label{fig:comp_network}
\end{figure*}

Where the experiments on RQ1 focus on the performances of different component combinations, is the focus of RQ2 towards comparing the best-performing Network-based RL agent (with Test Case Failure reward) with other test case prioritization methods.
\autoref{fig:comp_network} shows the results of the comparison against the three methods on each of the three data sets.
A comparison is made for every 30 CI cycles on the difference of the average NAPFD values of each cycle.
Positive differences show better performance by the comparison method, a negative difference shows better performance by \retecs{}.

During early CI cycles, the deterministic comparison methods show mostly better performance.
This corresponds to the initial exploration phase, where \retecs{} adapts to its environment.
After approximately 60 CI cycles, for Paint Control, it is able to prioritize with similar or better performance than the comparison methods.
Similar results are visible on the other two data sets, with a longer adaptation phase but less performance differences on IOF/ROL and an early comparable performance on GSDTSR.

For IOF/ROL, where the previous evaluation (see \autoref{fig:res_rewardfuns}) showed lower performance compared to Paint Control, also the comparison methods are not able to correctly prioritize failing test cases higher, as the small performance gap indicates.

For GSDTSR, \retecs{} is performing overall comparable with an NAPFD difference up to 0.2.
Due to the few failures within the data set, the exploration phase does not impact the performance in the early cycles as strongly as for the other two data sets.
Also, it appears as if the indicators for failing test cases are not as correlated to the previous test execution results as they were in the other data sets, which is visible from the comparatively low performance of the deterministic methods.

In summary, the results for RQ2 show, that \retecs{} can, starting from a model-free memory without initial knowledge about test case prioritization, in around 60 cycles, which corresponds to two month for daily intervals, learn to effectively prioritize test cases.
Its performance compares to that of basic deterministic test case prioritization methods.
For CI, this means that \retecs{} is a promising method for test case prioritization which adapts to environment specific indication of system failures.

\subsubsection{Internal Evaluation: Schedule Time Influence} \label{sec:sched_time_ratio}
In the experimental setup, the time limit for each CI cycle's reduced test schedule is set to 50\% of the execution time of the overall test suite $\mathcal{T}_i$.
To see how this choice influences the results and how it affects the learning process, an additional experiment is conducted with varying scheduling time ratios.

\begin{figure}
  \centering
  \begin{adjustbox}{trim=0.5cm 0.5cm 0.5cm 0.3cm}
 \scalebox{0.95}{\input{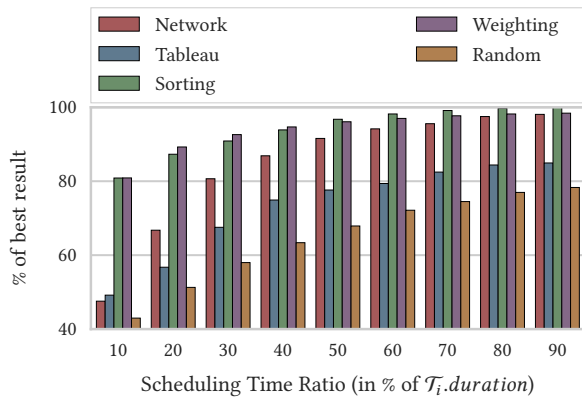}}%
  \end{adjustbox}
  \caption{Relative performance under different time limits. Shorter scheduling times reduce the information for rewards and delay learning. The performance differences for Network and Tableau also arise from the initial exploration phase, as shown in \autoref{fig:comp_network} (Data set: ABB Paint Control).}
  \label{fig:scheduling_time}%
\end{figure}

\autoref{fig:scheduling_time} shows the results on the Paint Control data set.
The NAPFD result is averaged over all CI cycles, which explains the overall better performance by the comparison methods due to an initial learning period.
As it is expected, performance decreases with lower time limits for all methods.
However, for RL agents a decreased scheduling time directly decreases available information for learning as fewer test cases can be executed and fewer actions can meaningfully be rewarded, resulting in a slower learning process.

Nevertheless, the decrease in performance is not directly proportional to the decrease in scheduling time, a sign that \retecs{} learns at some point how to prioritize test cases even though the amount of data in previous cycles was limited.

\subsection{Threats to Validity} \label{sec:threats}
\noindent\textbf{Internal.} % Why might we have reported wrong or incomplete results?
The first threat to internal validity is the influence of random decisions on the results.
To mitigate the threat, we repeated our experiments 30 times and report averaged results.

Another threat is related to the existence of faults within our implementation.
We approached this threat by applying established components, such as scikit-learn, within our software where appropriate.
Furthermore, our implementation is available online for inspection and reproduction of experiments.

Finally, many machine learning algorithms are sensible to their parameters and a feasible parameter set for one problem environment might not work for as well for different one.
During our experiments, the initially selected parameters were not changed for different problems to allow better comparison.
In a real-world setting, those parameters can be adjusted to tune the approach for the specific environment.

\noindent\textbf{External.} % Why might our approach not work for other people? Limit the generalization of the experiment to new case study applications
Our evaluation is based on data from three industrial data sets, which is a limitation regarding the wide variety of CI environments and failure distributions.
One of these data sets is publicly available, but according to our knowledge it has only been used in one publication and a different setting \cite{Elbaum2014}.
From what we have analyzed, there are no further public data sets available which include the required data, especially test verdicts over time.
This threat has to be addressed by additional experiments in different settings once further data is accessible.
To improve the data availability, we publish the other two data sets used in our experiments.

\noindent\textbf{Construct.} % Why is our approach stupid?
A threats to construct validity is the assumption, that each failed test cases indicates a different failure in the system under test.
This is not always true.
One test case can fail due to multiple failures in the system and one failure can lead to multiple failing test cases.
Based on the abstraction level of our method, this information is not easily available.
Nevertheless, our approach tries to find all failing test cases and thereby indirectly also all detectable failures.
To address the threat, we propose to include failure causes as input features in future work.

Further regarding the input features, our proposed method uses only few test case metadata to prioritize test cases and to reason about their importance for the test schedule.
In practical environments, more information about test cases or the system under test is available and should be utilized.

We compared our method to baseline approaches, but we have not considered additional techniques.
Although further methods exist in literature, they do not report results on comparable data sets or would need adjustment for our CI setting.

\subsection{Extensions} \label{sec:extensions}
The presented results give perspectives to extensions from two angles. 
First perspective is on the technical RL approach.
Through a pre-training phase the agent can internalize test case prioritization knowledge before actually prioritizing test cases and thereby improve the initial performance.
This can be approached by imitation of other methods \cite{Abbeel2004}, e.g. deterministic methods with desirable behavior, or by using historical data before it is introduced in the CI process \cite{Riedmiller2005}.
The second perspective focuses on the domain-specific approach of test case prioritization and selection.
Here, only few metadata of a test case and its history is facilitated.
The number of features of a test case should be extended to allow better reasoning of expected failures, e.g. links between source code changes and relevant test cases.
By including failure causes, scheduling of redundant test cases can be avoided and the effectiveness improved.

Furthermore, this work used a linear scheduling model, but in industrial environments more complex environments are encountered, e.g. multiple systems for test executions or additional constraints on test execution besides time limits.
Another extension of this work is therefore to integrate different scheduling methods under consideration of prioritization information and integration into the learning process \cite{Qu2008a}.

%%% Local Variables:
%%% mode: latex
%%% TeX-master: "../SGM17"
%%% End:
     % 
\section{Related Work}
\label{sec:related_work}
%\noindent
%This work relates to two main areas: 1) test case prioritization and selection for regression testing, and 2) machine learning for software testing. 
%
\noindent \textbf{Test case prioritization and selection for regression testing}: Previous work focuses on optimizing regression testing based on mainly three aspects: cost, coverage, and fault detection, or their combinations. In \cite{Siavash} authors propose an approach for test case selection and prioritization using the combination of Integer Linear Programming (ILP) and greedy methods by optimizing multiple criteria. Another study investigates coverage-based regression testing \cite{Nardo}, using four common prioritization techniques: a test selection technique, a test suite minimization technique and a hybrid approach that combines selection and minimization.
Similar approaches have been proposed using search-based algorithms \cite{Yu,Souza1}, including swarm optimization \cite{Souza} and ant colony optimization \cite{Noguchi}. Walcott \etal{} use genetic algorithms for time-aware regression test suite prioritization for frequent code rebuilding \cite{Walcott2006}.
Similarly, Zhang \etal{} propose time-aware prioritization using ILP \cite{Zhang2009}. Strandberg \etal{} \cite{Strandberg2016} apply a novel prioritization method with multiple factors in a real-world embedded software and show the improvement over industry practice.
Other regression test selection techniques have been proposed based on historical test data \cite{Marijan2013, Kim2002, Noor2015, Park}, code dependencies \cite{Gligoric}, or information retrieval \cite{Kwon2014,Saha2015}.
Despite various approaches to test optimization for regression testing, the challenge of applying most of them in practice lies in their complexity and the computational overhead typically required to collect and analyze different test parameters needed for prioritization, such as age, test coverage, etc.
By contrast, our approach based on RL is a lightweight method, which only uses historical results and its experience from previous CI cycles.
Furthermore, \retecs{} is adaptive and suited for dynamic environments with frequent changes in code and testing, and evolving test suites.

\noindent \textbf{Machine learning for software testing}:
Machine learning algorithms receive increasing attention in the context of software testing.
The work closest to ours is \cite{Busjaeger2016}, where Busjaeger and Xie use machine learning and multiple heuristic techniques to prioritize test cases in an industrial setting.
By combining various data sources and learning to rank in an agnostic way, this work makes a strong step into the definition of a general framework to automatically learn to rank test cases.
Our approach, only based on RL and ANN, takes another direction by providing a lightweight learning method using one source of data, namely test case failure history. 
Chen \etal{} \cite{Chen} uses semi-supervised clustering for regression test selection. The downside of such an approach may be higher computational complexity.
Other approaches include active learning for test classification \cite{Bowring}, combining machine learning and program slicing for regression test case prioritization \cite{Wang}, learning agent-based test case prioritization \cite{Abele}, or clustering approaches \cite{Chaurasia}.
RL has been previously used in combination with adaptation-based programming (ABP) for automated testing of software APIs, where the combination of RL and ABP successively selects calls to the API with the goal to increase test coverage, by \citeauthor{Groce2012} \cite{Groce2012}.
Furthermore, \citeauthor{Reichstaller2016} \cite{Reichstaller2016} apply RL to generate test cases for risk-based interoperability testing. Based on a model of the system under test, RL agents are trained to interact in an error-provoking way, i.e. they are encouraged to exploit possible interactions between components. Veanes \etal{} use RL for online formal testing of communication systems \cite{Veanes2006}. 
Based on the idea to see testing as a two-player game, RL is used to strengthen the tester's behavior when system and test cases are modeled as Input-Output Labeled Transition Systems. 
While this approach is appealing, \retecs{} applies RL for a completely different purpose, namely test case prioritization and selection.
Our approach aims at CI environments, which are characterized by strict time and effort constraints. 

%\subsection{Test Case Selection, Regression Testing, Test Suite Optimization}
%\cite{Busjaeger2016}: Learning for Test Prioritization: An Industrial Case Study
%\cite{Marijan2013}: Test Case Prioritization for Continuous Regression Testing: An Industrial Case Study
%\cite{Kwon2014}: Test case prioritization based on information retrieval concepts
%\cite{Walcott2006}, \cite{Zhang2009}: Time-Aware Test Suite Prioritization

%%% Local Variables:
%%% mode: latex
%%% TeX-master: "../SGM17"
%%% End:
          % 
\section{Conclusion}
\label{sec:conclusion}
\noindent
We presented \retecs{}, a novel lightweight method for test case prioritization and selection in Continuous Integration, combining reinforcement learning methods and historical test information.
\retecs{} is adaptive and learns important indicators for failing test cases during its runtime by observing test cases, test results, and its own actions and their effects.

Evaluation results show fast learning and adaptation of \retecs{} in three industrial case studies.
An effective prioritization strategy is discovered with a performance comparable to basic deterministic prioritization methods after an initial learning phase of approximately 60 CI cycles without previous training on test case prioritization.
Necessary domain knowledge is only reflected in a reward function to evaluate previous schedules.
The method is model-free, language-agnostic and requires no source code or program access.
It only requires test metadata, namely historical results, durations and last execution times.
However, we expect additional metadata to enhance the method's performance.

In our evaluation we compare different variants of RL agents for the ATCS problem.
Agents based on artificial neural networks have shown to be best performing, especially when trained with test case-individual reward functions.
While we applied only small networks in this work, with extended available data amounts, an extension towards larger networks and deep learning techniques can be a promising path for future research.

%%% Local Variables:
%%% mode: latex
%%% TeX-master: "../SGM17"
%%% End:
       % 

\balance
\bibliographystyle{ACM-Reference-Format}
\bibliography{refs}


\begin{thebibliography}{00}

%%% the Web address.

\ifx \showCODEN    \undefined \def \showCODEN     #1{\unskip}     \fi
\ifx \showDOI      \undefined \def \showDOI       #1{#1}\fi
\ifx \showISBNx    \undefined \def \showISBNx     #1{\unskip}     \fi
\ifx \showISBNxiii \undefined \def \showISBNxiii  #1{\unskip}     \fi
\ifx \showISSN     \undefined \def \showISSN      #1{\unskip}     \fi
\ifx \showLCCN     \undefined \def \showLCCN      #1{\unskip}     \fi
\ifx \shownote     \undefined \def \shownote      #1{#1}          \fi
\ifx \showarticletitle \undefined \def \showarticletitle #1{#1}   \fi
\ifx \showURL      \undefined \def \showURL       {\relax}        \fi
% invisible to TeX
\providecommand\bibfield[2]{#2}
\providecommand\bibinfo[2]{#2}
\providecommand\natexlab[1]{#1}
\providecommand\showeprint[2][]{arXiv:#2}

\bibitem[\protect\citeauthoryear{Abbeel and Ng}{Abbeel and Ng}{2004}]%
        {Abbeel2004}
\bibfield{author}{\bibinfo{person}{Pieter Abbeel} {and}
  \bibinfo{person}{Andrew~Y Ng}.} \bibinfo{year}{2004}\natexlab{}.
\newblock \showarticletitle{{Apprenticeship learning via inverse reinforcement
  learning}}.
\newblock \bibinfo{journal}{{\em Proceedings of the 21st International
  Conference on Machine Learning (ICML)\/}} (\bibinfo{year}{2004}),
  \bibinfo{pages}{1--8}.
\newblock
\showISBNx{1581138385 (ISBN)}
\showDOI{%
\url{https://doi.org/10.1145/1015330.1015430}}
\showeprint[arxiv]{1206.5264}


\bibitem[\protect\citeauthoryear{Abele and G{\"{o}}hner}{Abele and
  G{\"{o}}hner}{2014}]%
        {Abele}
\bibfield{author}{\bibinfo{person}{Sebastian Abele} {and}
  \bibinfo{person}{Peter G{\"{o}}hner}.} \bibinfo{year}{2014}\natexlab{}.
\newblock \showarticletitle{{Improving Proceeding Test Case Prioritization with
  Learning Software Agents}}. In \bibinfo{booktitle}{{\em Proceedings of the
  6th International Conference on Agents and Artificial Intelligence - Volume 2
  (ICAART)}}. \bibinfo{pages}{293--298}.
\newblock
\showISBNx{978-989-758-016-1}


\bibitem[\protect\citeauthoryear{Bowring, Rehg, and Harrold}{Bowring
  et~al\mbox{.}}{2004}]%
        {Bowring}
\bibfield{author}{\bibinfo{person}{James~F Bowring}, \bibinfo{person}{James~M
  Rehg}, {and} \bibinfo{person}{Mary~Jean Harrold}.}
  \bibinfo{year}{2004}\natexlab{}.
\newblock \showarticletitle{{Active Learning for Automatic Classification of
  Software Behavior}}. In \bibinfo{booktitle}{{\em Proceedings of the 2004 ACM
  SIGSOFT International Symposium on Software Testing and Analysis}} {\em
  (\bibinfo{series}{ISSTA '04})}. \bibinfo{publisher}{ACM},
  \bibinfo{address}{New York, NY, USA}, \bibinfo{pages}{195--205}.
\newblock
\showISBNx{1-58113-820-2}
\showDOI{%
\url{https://doi.org/10.1145/1007512.1007539}}


\bibitem[\protect\citeauthoryear{Busjaeger and Xie}{Busjaeger and Xie}{2016}]%
        {Busjaeger2016}
\bibfield{author}{\bibinfo{person}{Benjamin Busjaeger} {and}
  \bibinfo{person}{Tao Xie}.} \bibinfo{year}{2016}\natexlab{}.
\newblock \showarticletitle{{Learning for Test Prioritization: An Industrial
  Case Study}}. In \bibinfo{booktitle}{{\em Proceedings of the 2016 24th ACM
  SIGSOFT International Symposium on Foundations of Software Engineering}}.
  \bibinfo{publisher}{ACM}, \bibinfo{address}{New York, NY, USA},
  \bibinfo{pages}{975--980}.
\newblock
\showISBNx{1450342183}
\showDOI{%
\url{https://doi.org/10.1145/2950290.2983954}}


\bibitem[\protect\citeauthoryear{Chaurasia, Agarwal, and Gautam}{Chaurasia
  et~al\mbox{.}}{2015}]%
        {Chaurasia}
\bibfield{author}{\bibinfo{person}{G Chaurasia}, \bibinfo{person}{S Agarwal},
  {and} \bibinfo{person}{S~S Gautam}.} \bibinfo{year}{2015}\natexlab{}.
\newblock \showarticletitle{{Clustering based novel test case prioritization
  technique}}. In \bibinfo{booktitle}{{\em 2015 IEEE Students Conference on
  Engineering and Systems (SCES)}}. \bibinfo{publisher}{IEEE},
  \bibinfo{pages}{1--5}.
\newblock
\showDOI{%
\url{https://doi.org/10.1109/SCES.2015.7506447}}


\bibitem[\protect\citeauthoryear{Chen, Chen, Zhao, Xu, and Feng}{Chen
  et~al\mbox{.}}{2011}]%
        {Chen}
\bibfield{author}{\bibinfo{person}{S Chen}, \bibinfo{person}{Z Chen},
  \bibinfo{person}{Z Zhao}, \bibinfo{person}{B Xu}, {and} \bibinfo{person}{Y
  Feng}.} \bibinfo{year}{2011}\natexlab{}.
\newblock \showarticletitle{{Using semi-supervised clustering to improve
  regression test selection techniques}}. In \bibinfo{booktitle}{{\em 2011
  Fourth IEEE International Conference on Software Testing, Verification and
  Validation}}. \bibinfo{publisher}{IEEE}, \bibinfo{pages}{1--10}.
\newblock
\showISSN{2159-4848}
\showDOI{%
\url{https://doi.org/10.1109/ICST.2011.38}}


\bibitem[\protect\citeauthoryear{de~Souza, de~Miranda, Prudencio, and
  Barros}{de~Souza et~al\mbox{.}}{2011}]%
        {Souza1}
\bibfield{author}{\bibinfo{person}{Luciano~S de Souza},
  \bibinfo{person}{Pericles~BC de Miranda}, \bibinfo{person}{Ricardo~BC
  Prudencio}, {and} \bibinfo{person}{Flavia de~A Barros}.}
  \bibinfo{year}{2011}\natexlab{}.
\newblock \showarticletitle{{A Multi-objective Particle Swarm Optimization for
  Test Case Selection Based on Functional Requirements Coverage and Execution
  Effort}}. In \bibinfo{booktitle}{{\em 2011 IEEE 23rd International Conference
  on Tools with Artificial Intelligence}}. \bibinfo{publisher}{IEEE},
  \bibinfo{pages}{245--252}.
\newblock
\showISSN{1082-3409}
\showDOI{%
\url{https://doi.org/10.1109/ICTAI.2011.45}}


\bibitem[\protect\citeauthoryear{de~Souza, Prud{\^{e}}ncio, {de A. Barros}, and
  {da S. Aranha}}{de~Souza et~al\mbox{.}}{2013}]%
        {Souza}
\bibfield{author}{\bibinfo{person}{Luciano~S de Souza},
  \bibinfo{person}{Ricardo B~C Prud{\^{e}}ncio}, \bibinfo{person}{Flavia {de A.
  Barros}}, {and} \bibinfo{person}{Eduardo~H {da S. Aranha}}.}
  \bibinfo{year}{2013}\natexlab{}.
\newblock \showarticletitle{{Search based constrained test case selection using
  execution effort}}.
\newblock \bibinfo{journal}{{\em Expert Systems with Applications\/}}
  \bibinfo{volume}{40}, \bibinfo{number}{12} (\bibinfo{year}{2013}),
  \bibinfo{pages}{4887--4896}.
\newblock
\showISSN{0957-4174}
\showDOI{%
\url{https://doi.org/10.1016/j.eswa.2013.02.018}}


\bibitem[\protect\citeauthoryear{{Di Nardo}, Alshahwan, Briand, and
  Labiche}{{Di Nardo} et~al\mbox{.}}{2015}]%
        {Nardo}
\bibfield{author}{\bibinfo{person}{Daniel {Di Nardo}}, \bibinfo{person}{Nadia
  Alshahwan}, \bibinfo{person}{Lionel Briand}, {and} \bibinfo{person}{Yvan
  Labiche}.} \bibinfo{year}{2015}\natexlab{}.
\newblock \showarticletitle{{Coverage-based regression test case selection,
  minimization and prioritization: a case study on an industrial system}}.
\newblock \bibinfo{journal}{{\em Software Testing, Verification and
  Reliability\/}} \bibinfo{volume}{25}, \bibinfo{number}{4}
  (\bibinfo{year}{2015}), \bibinfo{pages}{371--396}.
\newblock
\showISSN{1099-1689}
\showDOI{%
\url{https://doi.org/10.1002/stvr.1572}}


\bibitem[\protect\citeauthoryear{Duvall, Matyas, and Glover}{Duvall
  et~al\mbox{.}}{2007}]%
        {duvall2007continuous}
\bibfield{author}{\bibinfo{person}{P~M Duvall}, \bibinfo{person}{S Matyas},
  {and} \bibinfo{person}{A Glover}.} \bibinfo{year}{2007}\natexlab{}.
\newblock \bibinfo{booktitle}{{\em {Continuous Integration: Improving Software
  Quality and Reducing Risk}}}.
\newblock \bibinfo{publisher}{Pearson Education}.
\newblock


\bibitem[\protect\citeauthoryear{Elbaum, Mclaughlin, and Penix}{Elbaum
  et~al\mbox{.}}{2014a}]%
        {Elbaum2014dataset}
\bibfield{author}{\bibinfo{person}{Sebastian Elbaum}, \bibinfo{person}{Andrew
  Mclaughlin}, {and} \bibinfo{person}{John Penix}.}
  \bibinfo{year}{2014}\natexlab{a}.
\newblock \bibinfo{title}{{The Google Dataset of Testing Results}}.
\newblock   (\bibinfo{year}{2014}).
\newblock
\showURL{%
\url{https://code.google.com/p/google-shared-dataset-of-test-suite-results/}}


\bibitem[\protect\citeauthoryear{Elbaum, Rothermel, and Penix}{Elbaum
  et~al\mbox{.}}{2014b}]%
        {Elbaum2014}
\bibfield{author}{\bibinfo{person}{Sebastian Elbaum}, \bibinfo{person}{Gregg
  Rothermel}, {and} \bibinfo{person}{John Penix}.}
  \bibinfo{year}{2014}\natexlab{b}.
\newblock \showarticletitle{{Techniques for improving regression testing in
  continuous integration development environments}}. In
  \bibinfo{booktitle}{{\em Proceedings of the 22nd ACM SIGSOFT International
  Symposium on Foundations of Software Engineering}}. \bibinfo{publisher}{ACM},
  \bibinfo{pages}{235--245}.
\newblock
\showISBNx{9781450330565}
\showDOI{%
\url{https://doi.org/10.1145/2635868.2635910}}


\bibitem[\protect\citeauthoryear{Fowler and Foemmel}{Fowler and
  Foemmel}{2006}]%
        {fowler2006continuous}
\bibfield{author}{\bibinfo{person}{Martin Fowler} {and} \bibinfo{person}{M
  Foemmel}.} \bibinfo{year}{2006}\natexlab{}.
\newblock \bibinfo{title}{{Continuous integration}}.
\newblock   (\bibinfo{year}{2006}).
\newblock
\showURL{%
\url{http://martinfowler.com/articles/continuousIntegration.html}}


\bibitem[\protect\citeauthoryear{Gligoric, Eloussi, and Marinov}{Gligoric
  et~al\mbox{.}}{2015}]%
        {Gligoric}
\bibfield{author}{\bibinfo{person}{M Gligoric}, \bibinfo{person}{L Eloussi},
  {and} \bibinfo{person}{D Marinov}.} \bibinfo{year}{2015}\natexlab{}.
\newblock \showarticletitle{{Ekstazi: Lightweight Test Selection}}. In
  \bibinfo{booktitle}{{\em Proceedings of the 37th International Conference on
  Software Engineering}}, Vol.~\bibinfo{volume}{2}. \bibinfo{pages}{713--716}.
\newblock
\showISSN{0270-5257}
\showDOI{%
\url{https://doi.org/10.1109/ICSE.2015.230}}


\bibitem[\protect\citeauthoryear{Groce, Fern, Pinto, Bauer, Alipour, Erwig, and
  Lopez}{Groce et~al\mbox{.}}{2012}]%
        {Groce2012}
\bibfield{author}{\bibinfo{person}{A. Groce}, \bibinfo{person}{A. Fern},
  \bibinfo{person}{J. Pinto}, \bibinfo{person}{T. Bauer}, \bibinfo{person}{A.
  Alipour}, \bibinfo{person}{M. Erwig}, {and} \bibinfo{person}{C. Lopez}.}
  \bibinfo{year}{2012}\natexlab{}.
\newblock \showarticletitle{Lightweight Automated Testing with Adaptation-Based
  Programming}. In \bibinfo{booktitle}{{\em 2012 IEEE 23rd International
  Symposium on Software Reliability Engineering}}. \bibinfo{pages}{161--170}.
\newblock
\showISSN{1071-9458}
\showDOI{%
\url{https://doi.org/10.1109/ISSRE.2012.1}}


\bibitem[\protect\citeauthoryear{Kim and Porter}{Kim and Porter}{2002}]%
        {Kim2002}
\bibfield{author}{\bibinfo{person}{Jung-Min Kim Jung-Min Kim} {and}
  \bibinfo{person}{A. Porter}.} \bibinfo{year}{2002}\natexlab{}.
\newblock \showarticletitle{{A history-based test prioritization technique for
  regression testing in resource constrained environments}}. In
  \bibinfo{booktitle}{{\em Proceedings of the 24th international conference on
  software engineering}}. \bibinfo{pages}{119--129}.
\newblock
\showDOI{%
\url{https://doi.org/10.1109/ICSE.2002.1007961}}


\bibitem[\protect\citeauthoryear{Kwon, Ko, Rothermel, and Staats}{Kwon
  et~al\mbox{.}}{2014}]%
        {Kwon2014}
\bibfield{author}{\bibinfo{person}{Jung-Hyun Kwon}, \bibinfo{person}{In-Young
  Ko}, \bibinfo{person}{Gregg Rothermel}, {and} \bibinfo{person}{Matt Staats}.}
  \bibinfo{year}{2014}\natexlab{}.
\newblock \showarticletitle{{Test case prioritization based on information
  retrieval concepts}}.
\newblock \bibinfo{journal}{{\em 2014 21st Asia-Pacific Software Engineering
  Conference (APSEC)\/}}  \bibinfo{volume}{1} (\bibinfo{year}{2014}),
  \bibinfo{pages}{19--26}.
\newblock
\showISBNx{9781479974252}
\showISSN{15301362}
\showDOI{%
\url{https://doi.org/10.1109/APSEC.2014.12}}


\bibitem[\protect\citeauthoryear{Lin}{Lin}{1992}]%
        {Lin1992}
\bibfield{author}{\bibinfo{person}{Long-Ji Lin}.}
  \bibinfo{year}{1992}\natexlab{}.
\newblock \showarticletitle{{Self-Improving Reactive Agents Based on
  Reinforcement Learning, Planning and Teaching}}.
\newblock \bibinfo{journal}{{\em Machine Learning\/}} \bibinfo{volume}{8},
  \bibinfo{number}{3-4} (\bibinfo{year}{1992}), \bibinfo{pages}{293--321}.
\newblock
\showISSN{15730565}
\showDOI{%
\url{https://doi.org/10.1023/A:1022628806385}}


\bibitem[\protect\citeauthoryear{Marijan, Gotlieb, and Sen}{Marijan
  et~al\mbox{.}}{2013}]%
        {Marijan2013}
\bibfield{author}{\bibinfo{person}{Dusica Marijan}, \bibinfo{person}{Arnaud
  Gotlieb}, {and} \bibinfo{person}{Sagar Sen}.}
  \bibinfo{year}{2013}\natexlab{}.
\newblock \showarticletitle{{Test case prioritization for continuous regression
  testing: An industrial case study}}. In \bibinfo{booktitle}{{\em 2013 29th
  IEEE International Conference on Software Maintenance (ICSM)}}.
  \bibinfo{pages}{540--543}.
\newblock
\showDOI{%
\url{https://doi.org/10.1109/ICSM.2013.91}}


\bibitem[\protect\citeauthoryear{Matari{\'{c}}}{Matari{\'{c}}}{1994}]%
        {Mataric1994}
\bibfield{author}{\bibinfo{person}{Maja~J Matari{\'{c}}}.}
  \bibinfo{year}{1994}\natexlab{}.
\newblock \showarticletitle{{Reward functions for accelerated learning}}. In
  \bibinfo{booktitle}{{\em Machine Learning: Proceedings of the Eleventh
  international conference}}. \bibinfo{pages}{181--189}.
\newblock
\showDOI{%
\url{https://doi.org/10.1.1.42.4313}}


\bibitem[\protect\citeauthoryear{Mirarab, {Akhlaghi Esfahani}, and
  Tahvildari}{Mirarab et~al\mbox{.}}{2012}]%
        {Siavash}
\bibfield{author}{\bibinfo{person}{Siavash Mirarab}, \bibinfo{person}{Soroush
  {Akhlaghi Esfahani}}, {and} \bibinfo{person}{Ladan Tahvildari}.}
  \bibinfo{year}{2012}\natexlab{}.
\newblock \showarticletitle{{Size-Constrained Regression Test Case Selection
  Using Multicriteria Optimization}}.
\newblock \bibinfo{journal}{{\em IEEE Transactions on Software Engineering\/}}
  \bibinfo{volume}{38}, \bibinfo{number}{4} (\bibinfo{date}{jul}
  \bibinfo{year}{2012}), \bibinfo{pages}{936--956}.
\newblock
\showISSN{0098-5589}
\showDOI{%
\url{https://doi.org/10.1109/TSE.2011.56}}


\bibitem[\protect\citeauthoryear{Noguchi, Washizaki, Fukazawa, Sato, and
  Ota}{Noguchi et~al\mbox{.}}{2015}]%
        {Noguchi}
\bibfield{author}{\bibinfo{person}{T Noguchi}, \bibinfo{person}{H Washizaki},
  \bibinfo{person}{Y Fukazawa}, \bibinfo{person}{A Sato}, {and}
  \bibinfo{person}{K Ota}.} \bibinfo{year}{2015}\natexlab{}.
\newblock \showarticletitle{{History-Based Test Case Prioritization for Black
  Box Testing Using Ant Colony Optimization}}. In \bibinfo{booktitle}{{\em 2015
  IEEE 8th International Conference on Software Testing, Verification and
  Validation (ICST)}}. \bibinfo{pages}{1--2}.
\newblock
\showISSN{2159-4848}
\showDOI{%
\url{https://doi.org/10.1109/ICST.2015.7102622}}


\bibitem[\protect\citeauthoryear{Noor and Hemmati}{Noor and Hemmati}{2015}]%
        {Noor2015}
\bibfield{author}{\bibinfo{person}{Tanzeem~Bin Noor} {and}
  \bibinfo{person}{Hadi Hemmati}.} \bibinfo{year}{2015}\natexlab{}.
\newblock \showarticletitle{{A similarity-based approach for test case
  prioritization using historical failure data}}.
\newblock \bibinfo{journal}{{\em 2015 IEEE 26th International Symposium on
  Software Reliability Engineering (ISSRE)\/}} (\bibinfo{year}{2015}),
  \bibinfo{pages}{58----68}.
\newblock
\showISBNx{7976931348623}


\bibitem[\protect\citeauthoryear{Orso and Rothermel}{Orso and
  Rothermel}{2014}]%
        {orso2014software}
\bibfield{author}{\bibinfo{person}{A Orso} {and} \bibinfo{person}{G
  Rothermel}.} \bibinfo{year}{2014}\natexlab{}.
\newblock \showarticletitle{{Software Testing: a Research Travelogue
  (2000--2014)}}. In \bibinfo{booktitle}{{\em Proceedings of the on Future of
  Software Engineering}}. \bibinfo{publisher}{ACM},
  \bibinfo{address}{Hyderabad, India}, \bibinfo{pages}{117--132}.
\newblock


\bibitem[\protect\citeauthoryear{Park, Ryu, and Baik}{Park
  et~al\mbox{.}}{2008}]%
        {Park}
\bibfield{author}{\bibinfo{person}{H Park}, \bibinfo{person}{H Ryu}, {and}
  \bibinfo{person}{J Baik}.} \bibinfo{year}{2008}\natexlab{}.
\newblock \showarticletitle{{Historical Value-Based Approach for Cost-Cognizant
  Test Case Prioritization to Improve the Effectiveness of Regression
  Testing}}. In \bibinfo{booktitle}{{\em 2008 Second International Conference
  on Secure System Integration and Reliability Improvement}}.
  \bibinfo{pages}{39--46}.
\newblock
\showDOI{%
\url{https://doi.org/10.1109/SSIRI.2008.52}}


\bibitem[\protect\citeauthoryear{Pedregosa, Varoquaux, Gramfort, Michel,
  Thirion, Grisel, Blondel, Prettenhofer, Weiss, Dubourg, Vanderplas, Passos,
  Cournapeau, Brucher, Perrot, and Duchesnay}{Pedregosa et~al\mbox{.}}{2011}]%
        {scikit-learn}
\bibfield{author}{\bibinfo{person}{F Pedregosa}, \bibinfo{person}{G Varoquaux},
  \bibinfo{person}{A Gramfort}, \bibinfo{person}{V Michel}, \bibinfo{person}{B
  Thirion}, \bibinfo{person}{O Grisel}, \bibinfo{person}{M Blondel},
  \bibinfo{person}{P Prettenhofer}, \bibinfo{person}{R Weiss},
  \bibinfo{person}{V Dubourg}, \bibinfo{person}{J Vanderplas},
  \bibinfo{person}{A Passos}, \bibinfo{person}{D Cournapeau},
  \bibinfo{person}{M Brucher}, \bibinfo{person}{M Perrot}, {and}
  \bibinfo{person}{E Duchesnay}.} \bibinfo{year}{2011}\natexlab{}.
\newblock \showarticletitle{{Scikit-learn: Machine Learning in
  {\{}P{\}}ython}}.
\newblock \bibinfo{journal}{{\em Journal of Machine Learning Research\/}}
  \bibinfo{volume}{12} (\bibinfo{year}{2011}), \bibinfo{pages}{2825--2830}.
\newblock


\bibitem[\protect\citeauthoryear{Qu, Nie, and Xu}{Qu et~al\mbox{.}}{2008}]%
        {Qu2008a}
\bibfield{author}{\bibinfo{person}{Bo Qu}, \bibinfo{person}{Changhai Nie},
  {and} \bibinfo{person}{Baowen Xu}.} \bibinfo{year}{2008}\natexlab{}.
\newblock \showarticletitle{{Test case prioritization for multiple processing
  queues}}.
\newblock \bibinfo{journal}{{\em 2008 International Symposium on Information
  Science and Engineering (ISISE)\/}}  \bibinfo{volume}{2}
  (\bibinfo{year}{2008}), \bibinfo{pages}{646--649}.
\newblock
\showISBNx{9780769534947}
\showISSN{2160-1283}
\showDOI{%
\url{https://doi.org/10.1109/ISISE.2008.106}}


\bibitem[\protect\citeauthoryear{Qu, Cohen, and Woolf}{Qu
  et~al\mbox{.}}{2007}]%
        {Qu2007}
\bibfield{author}{\bibinfo{person}{Xiao Qu}, \bibinfo{person}{Myra~B. Cohen},
  {and} \bibinfo{person}{Katherine~M. Woolf}.} \bibinfo{year}{2007}\natexlab{}.
\newblock \showarticletitle{{Combinatorial interaction regression testing: A
  study of test case generation and prioritization}}. In
  \bibinfo{booktitle}{{\em IEEE International Conference on Software
  Maintenance, 2007 (ICSM)}}. \bibinfo{publisher}{IEEE},
  \bibinfo{pages}{255--264}.
\newblock
\showISBNx{1424412560}
\showISSN{1063-6773}


\bibitem[\protect\citeauthoryear{Reichstaller, Eberhardinger, Knapp, Reif, and
  Gehlen}{Reichstaller et~al\mbox{.}}{2010}]%
        {Reichstaller2016}
\bibfield{author}{\bibinfo{person}{Andre~Andr{\'{e}} Reichstaller},
  \bibinfo{person}{Benedikt Eberhardinger}, \bibinfo{person}{Alexander Knapp},
  \bibinfo{person}{Wolfgang Reif}, {and} \bibinfo{person}{Marcel Gehlen}.}
  \bibinfo{year}{2010}\natexlab{}.
\newblock \showarticletitle{{Risk-Based Interoperability Testing Using
  Reinforcement Learning}}. In \bibinfo{booktitle}{{\em 28th IFIP WG 6.1
  International Conference, ICTSS 2016, Graz, Austria, October 17-19, 2016,
  Proceedings}}, \bibfield{editor}{\bibinfo{person}{Franz Wotawa},
  \bibinfo{person}{Mihai Nica}, {and} \bibinfo{person}{Natalia Kushik}} (Eds.),
  Vol.~\bibinfo{volume}{6435}. \bibinfo{publisher}{Springer International
  Publishing}, \bibinfo{address}{Cham}, \bibinfo{pages}{52--69}.
\newblock
\showISBNx{978-3-642-16572-6}
\showISSN{03029743}
\showDOI{%
\url{https://doi.org/10.1007/978-3-642-16573-3}}


\bibitem[\protect\citeauthoryear{Riedmiller}{Riedmiller}{2005}]%
        {Riedmiller2005}
\bibfield{author}{\bibinfo{person}{Martin Riedmiller}.}
  \bibinfo{year}{2005}\natexlab{}.
\newblock \showarticletitle{{Neural fitted Q iteration - First experiences with
  a data efficient neural Reinforcement Learning method}}. In
  \bibinfo{booktitle}{{\em European Conference on Machine Learning}}.
  \bibinfo{publisher}{Springer}, \bibinfo{pages}{317--328}.
\newblock
\showISBNx{3540292438}
\showISSN{03029743}
\showDOI{%
\url{https://doi.org/10.1007/11564096_32}}


\bibitem[\protect\citeauthoryear{Rothermel, Untch, Chu, and Harrold}{Rothermel
  et~al\mbox{.}}{1999}]%
        {rothermel1999}
\bibfield{author}{\bibinfo{person}{Gregg Rothermel}, \bibinfo{person}{Roland~H
  Untch}, \bibinfo{person}{Chengyun Chu}, {and} \bibinfo{person}{Mary~Jean
  Harrold}.} \bibinfo{year}{1999}\natexlab{}.
\newblock \showarticletitle{{Test case prioritization: An empirical study}}. In
  \bibinfo{booktitle}{{\em Software Maintenance, 1999.(ICSM'99) Proceedings.
  IEEE International Conference on}}. IEEE, \bibinfo{pages}{179--188}.
\newblock


\bibitem[\protect\citeauthoryear{Rothermel, Untch, Chu, Harrold, and
  Society}{Rothermel et~al\mbox{.}}{2001}]%
        {Rothermel2001}
\bibfield{author}{\bibinfo{person}{Gregg Rothermel}, \bibinfo{person}{Roland~H
  Untch}, \bibinfo{person}{Chengyun Chu}, \bibinfo{person}{Mary~Jean Harrold},
  {and} \bibinfo{person}{Ieee~Computer Society}.}
  \bibinfo{year}{2001}\natexlab{}.
\newblock \showarticletitle{{Prioritizing Test Cases For Regression Testing}}.
\newblock \bibinfo{journal}{{\em IEEE Transactions on Software Engineering\/}}
  \bibinfo{volume}{27}, \bibinfo{number}{10} (\bibinfo{year}{2001}),
  \bibinfo{pages}{929--948}.
\newblock
\showDOI{%
\url{https://doi.org/10.1145/347324.348910}}


\bibitem[\protect\citeauthoryear{Saha, Zhang, Khurshid, and Perry}{Saha
  et~al\mbox{.}}{2015}]%
        {Saha2015}
\bibfield{author}{\bibinfo{person}{Ripon~K Saha}, \bibinfo{person}{L Zhang},
  \bibinfo{person}{S Khurshid}, {and} \bibinfo{person}{D~E Perry}.}
  \bibinfo{year}{2015}\natexlab{}.
\newblock \showarticletitle{{An Information Retrieval Approach for Regression
  Test Prioritization Based on Program Changes}}. In \bibinfo{booktitle}{{\em
  Software Engineering (ICSE), 2015 IEEE/ACM 37th IEEE International Conference
  on}}, Vol.~\bibinfo{volume}{1}. \bibinfo{pages}{268--279}.
\newblock
\showISBNx{9781479919345}
\showISSN{0270-5257}
\showDOI{%
\url{https://doi.org/10.1109/ICSE.2015.47}}


\bibitem[\protect\citeauthoryear{Stolberg}{Stolberg}{2009}]%
        {stolberg2009enabling}
\bibfield{author}{\bibinfo{person}{S Stolberg}.}
  \bibinfo{year}{2009}\natexlab{}.
\newblock \showarticletitle{{Enabling agile testing through continuous
  integration}}. In \bibinfo{booktitle}{{\em Agile Conference, 2009.
  AGILE'09.}} IEEE, \bibinfo{pages}{369--374}.
\newblock


\bibitem[\protect\citeauthoryear{Strandberg, Sundmark, Afzal, Ostrand, and
  Weyuker}{Strandberg et~al\mbox{.}}{2016}]%
        {Strandberg2016}
\bibfield{author}{\bibinfo{person}{Per~Erik Strandberg},
  \bibinfo{person}{Daniel Sundmark}, \bibinfo{person}{Wasif Afzal},
  \bibinfo{person}{Thomas Ostrand}, {and} \bibinfo{person}{Elaine Weyuker}.}
  \bibinfo{year}{2016}\natexlab{}.
\newblock \showarticletitle{{Experience Report: Automated System Level
  Regression Test Prioritization Using Multiple Factors}}. In
  \bibinfo{booktitle}{{\em Software Reliability Engineering (ISSRE), 2016 IEEE
  27th International Symposium on}}. \bibinfo{publisher}{IEEE},
  \bibinfo{pages}{12----23}.
\newblock


\bibitem[\protect\citeauthoryear{Sutton and Barto}{Sutton and Barto}{1998}]%
        {Sutton1998}
\bibfield{author}{\bibinfo{person}{Richard~S. Sutton} {and}
  \bibinfo{person}{Andrew~G. Barto}.} \bibinfo{year}{1998}\natexlab{}.
\newblock \bibinfo{booktitle}{{\em {Reinforcement Learning: An Introduction}\/}
  (\bibinfo{edition}{1st} ed.)}.
\newblock \bibinfo{publisher}{MIT press Cambridge}.
\newblock
\showISBNx{0262193981}
\showISSN{1045-9227}
\showDOI{%
\url{https://doi.org/10.1109/TNN.1998.712192}}


\bibitem[\protect\citeauthoryear{{Van Hasselt} and Wiering}{{Van Hasselt} and
  Wiering}{2007}]%
        {VanHasselt2007}
\bibfield{author}{\bibinfo{person}{Hado {Van Hasselt}} {and}
  \bibinfo{person}{Marco~A Wiering}.} \bibinfo{year}{2007}\natexlab{}.
\newblock \showarticletitle{{Reinforcement learning in continuous action
  spaces}}.
\newblock \bibinfo{journal}{{\em Proceedings of the 2007 IEEE Symposium on
  Approximate Dynamic Programming and Reinforcement Learning, ADPRL 2007\/}}
  (\bibinfo{year}{2007}), \bibinfo{pages}{272--279}.
\newblock
\showISBNx{1424407060}
\showDOI{%
\url{https://doi.org/10.1109/ADPRL.2007.368199}}


\bibitem[\protect\citeauthoryear{{Van Rossum, Guido and Drake Jr}}{{Van Rossum,
  Guido and Drake Jr}}{1995}]%
        {Rossum:1995:PRM:869369}
\bibfield{author}{\bibinfo{person}{Fred~L {Van Rossum, Guido and Drake Jr}}.}
  \bibinfo{year}{1995}\natexlab{}.
\newblock \bibinfo{booktitle}{{\em {Python Reference Manual}}}.
\newblock \bibinfo{type}{{T}echnical {R}eport}. \bibinfo{address}{Amsterdam,
  The Netherlands, The Netherlands}.
\newblock


\bibitem[\protect\citeauthoryear{Veanes, Roy, and Campbell}{Veanes
  et~al\mbox{.}}{2006}]%
        {Veanes2006}
\bibfield{author}{\bibinfo{person}{Margus Veanes}, \bibinfo{person}{Pritam
  Roy}, {and} \bibinfo{person}{Colin Campbell}.}
  \bibinfo{year}{2006}\natexlab{}.
\newblock \bibinfo{booktitle}{{\em {Online Testing with Reinforcement
  Learning}}}.
\newblock \bibinfo{publisher}{Springer Berlin Heidelberg},
  \bibinfo{address}{Berlin, Heidelberg}, \bibinfo{pages}{240--253}.
\newblock
\showISBNx{978-3-540-49703-5}
\showDOI{%
\url{https://doi.org/10.1007/11940197_16}}


\bibitem[\protect\citeauthoryear{Walcott, Soffa, Kapfhammer, and Roos}{Walcott
  et~al\mbox{.}}{2006}]%
        {Walcott2006}
\bibfield{author}{\bibinfo{person}{K~R Walcott}, \bibinfo{person}{M~L Soffa},
  \bibinfo{person}{G~M Kapfhammer}, {and} \bibinfo{person}{R~S Roos}.}
  \bibinfo{year}{2006}\natexlab{}.
\newblock \showarticletitle{{Time-Aware Test Suite Prioritization}}. In
  \bibinfo{booktitle}{{\em Proceedings of the 2006 International Symposium on
  Software Testing and Analysis (ISSTA)}}. \bibinfo{publisher}{ACM},
  \bibinfo{address}{Portland, Maine, USA}, \bibinfo{pages}{1--12}.
\newblock


\bibitem[\protect\citeauthoryear{Wang, Yang, and Yang}{Wang
  et~al\mbox{.}}{2011}]%
        {Wang}
\bibfield{author}{\bibinfo{person}{Farn Wang}, \bibinfo{person}{Shun-Ching
  Yang}, {and} \bibinfo{person}{Ya-Lan Yang}.} \bibinfo{year}{2011}\natexlab{}.
\newblock \bibinfo{booktitle}{{\em {Regression Testing Based on Neural Networks
  and Program Slicing Techniques}}}.
\newblock \bibinfo{publisher}{Springer Berlin Heidelberg},
  \bibinfo{address}{Berlin, Heidelberg}, \bibinfo{pages}{409--418}.
\newblock
\showISBNx{978-3-642-25658-5}
\showDOI{%
\url{https://doi.org/10.1007/978-3-642-25658-5_50}}


\bibitem[\protect\citeauthoryear{Yu, Xu, and Tsai}{Yu et~al\mbox{.}}{2010}]%
        {Yu}
\bibfield{author}{\bibinfo{person}{Lian Yu}, \bibinfo{person}{Lei Xu}, {and}
  \bibinfo{person}{Wei-Tek Tsai}.} \bibinfo{year}{2010}\natexlab{}.
\newblock \bibinfo{booktitle}{{\em {Time-Constrained Test Selection for
  Regression Testing}}}.
\newblock \bibinfo{publisher}{Springer Berlin Heidelberg},
  \bibinfo{address}{Berlin, Heidelberg}, \bibinfo{pages}{221--232}.
\newblock
\showISBNx{978-3-642-17313-4}
\showDOI{%
\url{https://doi.org/10.1007/978-3-642-17313-4_23}}


\bibitem[\protect\citeauthoryear{Zhang, Hou, Guo, Xie, and Mei}{Zhang
  et~al\mbox{.}}{2009}]%
        {Zhang2009}
\bibfield{author}{\bibinfo{person}{Lu Zhang}, \bibinfo{person}{Shan-Shan Hou},
  \bibinfo{person}{Chao Guo}, \bibinfo{person}{Tao Xie}, {and}
  \bibinfo{person}{Hong Mei}.} \bibinfo{year}{2009}\natexlab{}.
\newblock \showarticletitle{{Time-aware test-case prioritization using integer
  linear programming}}.
\newblock \bibinfo{journal}{{\em Proceedings of the eighteenth International
  Symposium on Software Testing and Analysis (ISSTA)\/}}
  (\bibinfo{year}{2009}), \bibinfo{pages}{213--224}.
\newblock
\showISBNx{9781605583389}
\showDOI{%
\url{https://doi.org/10.1145/1572272.1572297}}


\bibitem[\protect\citeauthoryear{Zhang}{Zhang}{2004}]%
        {zhang2004}
\bibfield{author}{\bibinfo{person}{Tong Zhang}.}
  \bibinfo{year}{2004}\natexlab{}.
\newblock \showarticletitle{{Solving large scale linear prediction problems
  using stochastic gradient descent algorithms}}. In \bibinfo{booktitle}{{\em
  Proceedings of the twenty-first international conference on Machine
  learning}}. ACM, \bibinfo{pages}{116}.
\newblock


\end{thebibliography}
\end{document}